# Quantified Uncertainty in Thermodynamic Modeling for Materials Design

Noah H Paulson[1,*], Brandon J Bocklund[2], Richard A Otis[3], Zi-Kui Liu[2], Marius Stan[1]

[1]Applied Materials Division, Argonne National Laboratory, Argonne, IL 60439
[2]Department of Material Science and Engineering, Pennsylvania State University, University Park, PA 16802
[3]Engineering and Science Directorate, Jet Propulsion Laboratory, California Institute of Technology, Pasadena, CA 91109

## Abstract

Phase fractions, compositions and energies of the stable phases as a function of macroscopic composition, temperature, and pressure (X-T-P) are the principle correlations needed for the design of new materials and improvement of existing materials. They are the outcomes of thermodynamic modeling based on the CALculation of PHAse Diagrams (CALPHAD) approach. The accuracy of CALPHAD predictions vary widely in X-T-P space due to experimental error, model inadequacy and unequal data coverage. In response, researchers have developed frameworks to quantify the uncertainty of thermodynamic property model parameters and propagate it to phase diagram predictions. In previous studies, uncertainty was represented as intervals on phase boundaries (with respect to composition) or invariant reactions (with respect to temperature) and was unable to represent the uncertainty in eutectoid reactions or in the stability of phase regions. In this work, we propose a suite of tools that leverages samples from the multivariate model parameter distribution to represent uncertainty in forms that surpass previous limitations and are well suited to materials design. These representations include the distribution of phase diagrams and their features, as well as the dependence of phase stability and the distributions of phase fraction, composition activity and Gibbs energy on X-T-P location - irrespective of the total number of components. Most critically, the new methodology allows the material designer to interrogate a certain composition and temperature domain and get in return the probability of different phases to be stable, which can positively impact materials design.

* Corresponding author. Argonne National Laboratory, 9700 Cass Avenue, Lemont, IL 60439, USA. 630-252-1697
E-mail addresses: npaulson@anl.gov (N.H. Paulson), bocklund@psu.edu (B.J. Bocklund), Richard.Otis@jpl.nasa.gov (R.A. Otis), liu@matse.psu.edu (Z-K. Liu), mstan@anl.gov (M. Stan)

# 1 Introduction

Prediction of the stability of phases in thermodynamic equilibrium is a fundamental part of the design of multicomponent materials. The question of what phases are present, along with their percentages and their compositions at a particular overall composition, temperature and pressure (X-T-P), provides basic information regarding the suitability of a material for a desired application [1]. Even in the development of metastable materials, it is critical to have an accurate understanding of the equilibrium phase stability over X-T-P space to identify and refine optimal processing routes. Furthermore, these basic quantities serve as building blocks for more complex predictive methods - for example the prediction of the evolution of microstructures during processing (including diffusion and precipitation simulations)[2].

The CALPHAD approach has become the preferred method for the prediction of phase stability due to its predictive power in regions of X-T-P space with limited direct experimental or simulated information [1]. Instead of simply classifying phase space based on the observation of phase stability, the calibration of Gibbs energies modeled by Redlich-Kister polynomials within the compound energy formalism [3] in the CALPHAD approach enables physically reasonable extrapolations to regions without measurements and even into systems with higher numbers of components. This is especially necessary for multi-component systems where it becomes prohibitively expensive to experimentally sample phase stability in the X-T-P domain with sufficient coverage to justify the naïve approach.

CALPHAD predictions of phase stability are uncertain to a lesser or greater degree depending on the location in X-T-P space. Uncertainty in CALPHAD predictions derives from a number of sources, including both random and systematic errors in the measurement or simulation of quantities of interest used to calibrate the CALPHAD models, as well as the choice of specific model forms utilized to describe the thermodynamic properties of the phases [4], [5]. While infrequently addressed in a rigorous manner, uncertainty in CALPHAD predictions has implications for materials design efforts. Materials designers are well aware that a given CALPHAD database may represent certain X-T-P regions with low accuracy and in some cases not even reproduce the structure of the experimentally observed phase diagram. While experience can inform a qualitative understanding of the bounds of reliability of a prediction, the results of failed intuition are costly. For example, a CALPHAD equilibrium calculation may neglect the presence of a deleterious phase at an X-T-P location where the stability of that phase is not well represented. Without any indication of uncertainty, the CALPHAD result would not hint at the potential for the phase being present - resulting in a large loss of time and capital should the material be produced as a batch.

Over the past several decades, a number of authors have presented frameworks for uncertainty quantification (UQ) of CALPHAD models through both frequentist [6] and Bayesian [4], [5], [7]–[12] paradigms. In all known published works, uncertainty in the parameters of the thermodynamic models of each phase have been analytically propagated to the phase boundaries through the moments of the parameter distributions [12] or numerically through samples of the distributions [4], [11]. This representation of the uncertainty of predicted phase stabilities is limited in what questions it can address. Firstly, the intervals over these phase boundaries are constructed to quantify the uncertainty due to variation in either temperature, composition or pressure, but not a combination of these independent variables. It is not always clear which of these variables are selected, or if they will reasonably capture the uncertainty. Along similar lines, this approach does not explicitly address the intersection of multiple phase boundaries, nor the potential for phases to fall out of equilibrium for a subset of the distribution of model parameters. Finally, this representation cannot be easily extended to systems of three or more components.

In this work, we propose a number of methods that extend beyond quantifying the uncertainty of phase boundaries in directions well suited to materials design challenges. Each method leverages the distribution of CALPHAD model parameters that results from Monte Carlo optimization runs. We first present an approach to gain a qualitative understanding of the uncertainty in the phase diagram. We then demonstrate how to quantify the uncertainty of the location of invariant points. The most important result of this work for design applications is a method that polls the phase stability of an X-T-P point, irrespective of the number of components under consideration, and returns the probabilities that each phase is stable. This same approach results in probability distributions for phase fractions, compositions, activities, sublattice site fractions, Gibbs energies and all of the properties that are related to the first and second derivatives of the Gibbs energy. Furthermore, the techniques are trivially extended to metastable equilibria where uncertainty quantification is critical. These methodologies and their relevance to materials design are demonstrated through a case study with the Cu-Mg binary system utilizing CALPHAD model parameter samples obtained from MCMC optimization in ESPEI [5], [13].

# 2 Methodology

Our approach leverages Monte Carlo samples from the distribution of parameters for all models to quantify uncertainty of the phase diagram and phase stability. Each sample of this parameter distribution is a vector whose elements correspond to the model parameters. It follows that each sample also has a uniquely associated Gibbs energy description over the relevant X-T-P domain of the thermodynamic models. In other words, at every distinct X-T-P point, there exists a distribution of stable phases, phase fractions, compositions, and activities, along with any other calculable properties of interest.

An intuitive way to utilize the distribution of model parameters is to superimpose the binary or ternary phase diagrams resulting from each parameter set [14]. At first glance, this visualization may appear analogous to plotting the uncertainty intervals for each phase boundary; however, this approach captures all features of the phase diagram, including uncertainty in the location of invariant points and the presence or absence of features in the phase diagram including the stability of entire phases. Furthermore, it is not limited by the need to define uncertainty intervals based on a single state variable. While this approach provides valuable insight into binary phase diagram uncertainty, there still exists a need to provide quantitative uncertainties for an arbitrary number of state variables.

In certain design scenarios, it is critical to know the location of invariant points, including eutectoids, accurately. To address this need, we take the same approach as before and poll the X-T-P of the invariant of interest for samples from the model parameter distribution. Initially, we simply present scatter plots for the invariant locations in X-T-P space. It may also be desirable to construct uncertainty intervals for the invariant location to provide the probability of the invariant residing within certain regions. To this end, we employ kernel density estimation [15] to estimate the probability density function (PDF) of the sampled X-T-P locations and compute probability density contours labeled by the percent volume of the total distribution they contain. If necessary, the practitioner may arbitrarily delineate regions of X-T-P space and compute the probability of the invariant falling into it. For example, it is possible to estimate the probability that a eutectoid lies above some minimum operating temperature or other design parameter.

The most powerful application of this methodology is to compute uncertainties for any measure of interest and for any desired X-T-P point or region based on the values of those measures for the collection of sampled parameter values. This approach is advantageous as it frees the representation of uncertainty from the number of components in the materials system of interest. In other words, for each X-T-P point, the uncertainty in the phase stability (expressed as a probability) or any other property (typically expressed as probability distributions) are interpreted the same way whether it is a binary or 10 component system. As certain properties are phase specific (e.g. composition or activity), it is important to first mask those properties where the desired phase is not present before computing their probability distribution. Consequently, when visualizing the probability distribution for any quantity of interest it is important to note the total number of points used to advertise the statistical significance of the results.

## 3 Case Study
### 3.1 Implementation Details

The methodology discussed in the previous section is demonstrated for the Cu-Mg binary system. Bayesian inference for the CALPHAD model parameters is performed using the open source ESPEI [5] and pycalphad [16] Python packages. We assume that the reader has a working knowledge of Bayesian inference, and recommend our previous work alongside standard texts for an introduction and further detail [17], [18]. The selected model forms are identical to the authors' recent study highlighting the use of ESPEI for CALPHAD optimization [5]. Starting parameter values for MCMC posterior sampling are obtained by performing an initial (and deterministic) linear fitting of the Gibbs energy with thermochemical data, employing generic polynomials for temperature dependence and Redlich-Kister polynomials for compositional dependence. ESPEI provides a choice among uninformative (improper), uniform, Gaussian and triangular prior distributions. In this work, we observe that MCMC chains are likely to wander extensively in parameter space unless parameter values are reasonably bounded by the priors. This is expected since CALPHAD models are typically underdetermined when compared strictly with the data, and require expert input to maintain "physical reasonableness." Consequently, we choose triangular prior distributions with minimum and maximum parameter values and the center of mass corresponding to $p \pm 0.5p$ and $p$, respectively, where $p$ is the starting parameter value.

Within ESPEI, the Affine Invariant Ensemble Sampler [19] algorithm (via the emcee Python package [20]) is employed for MCMC Bayesian inference. In this algorithm, 150 MCMC walkers explore parameter space and optimally guide future walker movements. The walkers are initialized from a random sampling of univariate Gaussian distributions centered at $p$ with a $0.1p$ standard deviation. We run ESPEI until the walkers percolate a region in parameter space whose bounds do not appreciably change with increasing iterations for each parameter. In some cases, it is desirable to employ more qualitative measures of convergence such as the Gelman-Rubin statistic [17]; however, we do not employ this metric, as it would require us to perform several additional ESPEI runs (the walkers within a single run are dependent). The ESPEI runs performed in this work required 34 hours on a 36-core node on the Bebop cluster at Argonne National Laboratory to converge after 300 iterations including burn-in. After convergence, the walker locations in parameter space for the independent samples are employed for the uncertainty quantification approaches in the remainder of this section. Figure 1 below displays the convergence behavior of selected parameters for all walkers. The paths of two walkers in each subfigure are highlighted in black.

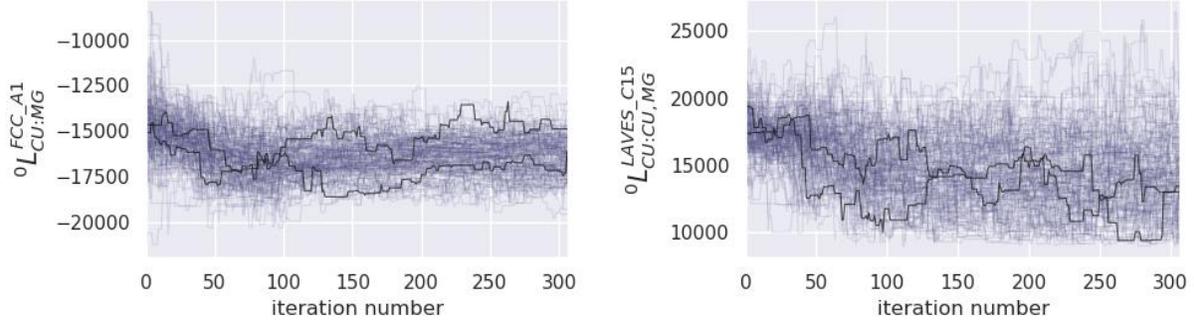

Figure 1: Parameter values for all 150 walkers are plotted versus iteration for parameters corresponding to the (a) FCC_A1 and (b) LAVES_C15 phases. In each figure, two chains are highlighted in black for the sake of visibility.

## 3.2 Results

The first analysis after the MCMC optimization is often focused on the equilibrium phase diagram. In Fig. 2, we superimpose the phase boundaries for the 150 parameter-sets corresponding to the final ESPEI iteration. This representation provides a general view of the phase diagram and the relative uncertainty of different features. For example, in Fig. 2a, the FCC_A1 boundary has a high degree of uncertainty compared to the CUMG2 or HCP_A3 phases. Let us suppose that the region surrounding the eutectic for FCC_A1, LAVES_C15 and LIQUID phases is of the greatest interest for design purposes. We can perform equilibrium calculations in a fine grid in this area to reduce the total computational expense and obtain a more refined view of the uncertainty. Figure 2b displays the superimposed phase boundaries with $x_{Mg}$ between 0.0 and 0.35 and temperature between 950K and 1100K.

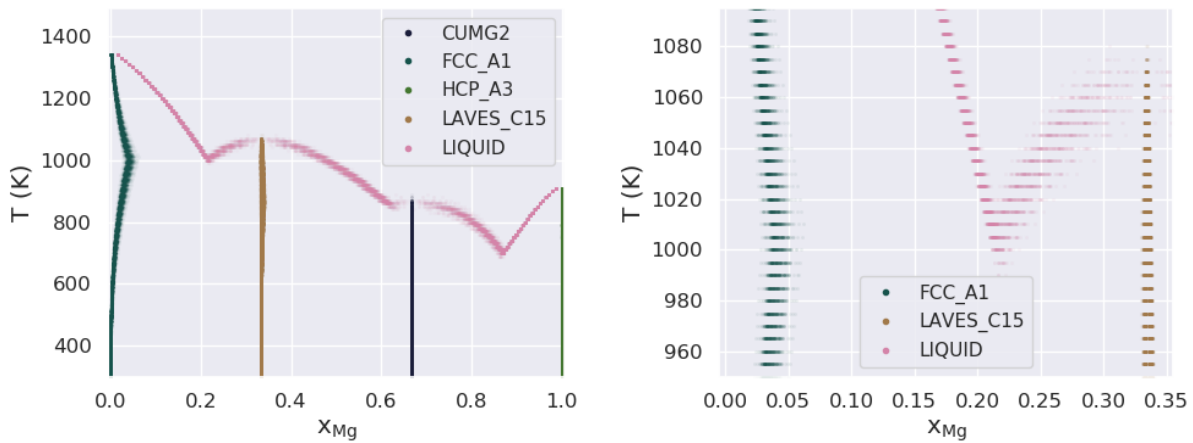

Figure 2: Phase boundaries for all 150 parameter sets are plotted on the same phase diagram with transparency (a) for all $x_{Mg}$ and T, and (b) for $x_{Mg}$ between 0.0 and 0.35 and T between 950 and 1100K.

It is also desirable to obtain quantitative information regarding the uncertainty in the location of the eutectoid. We produce Fig. 3 for this purpose, where we plot the location of the eutectoid for 750 model parameter vectors representing all 150 walkers for the last five ESPEI iterations. On this scatter plot, we superimpose the associated 68% and 95% uncertainty intervals (or Bayesian credible intervals containing 68 and 95 percent of the invariant samples respectively). This representation allows a materials designer to gauge the models' confidence in the location of the eutectoid. The designer may perform a similar analysis to determine that 6.8% of invariant samples have $x_{Mg}$ between 0.212 and 0.217, and temperatures between 999K and 1007K. This type of calculation can in fact be performed for any arbitrary region of X-T-P space, granting significant flexibility to the analysis.

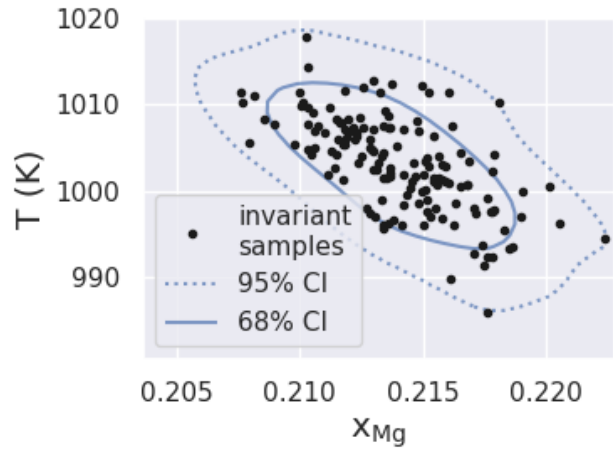

Figure 3: The FCC_A1 - LAVES_C15 – LIQUID eutectics are plotted for all 750 sampled parameter sets with 68% and 95% uncertainty intervals.

If we require further information about the phase stability and other properties at an X-T-P point near the eutectoid we can perform equilibrium calculations for all parameter vector samples at the mean eutectoid location ($x_{Mg}$: 0.214, T:1003). This results in the phase stability statistics presented in Table 1. For the sampled parameter vectors (750 corresponding to the last 5 ESPEI iterations), Table 1 shows that the FCC_A1+LAVES_C15 mixture is most common, being stable for over 50% of the samples, while LAVES_C15+LIQUID is least common with roughly a 4% probability (given the combination of model form and calibration data). We can also estimate the probability distributions for various quantities of interest, including the phase fraction, phase composition, activity and Gibbs energy from the same set of equilibrium calculations. Figure 4 presents these probability distributions as histograms for the FCC_A1 phase at the same X-T-P location for parameter sets where the FCC_A1+LAVES_C15 phase is stable.

By representing the probability of phase regions being stable at a particular concentration and temperature point in the diagram, we open the door to non-visual representations of uncertainty that can be extended to multi-components systems (ternary, quaternary, etc). The material designer can now interrogate a certain composition and temperature domain, and get in return the probability of different phases to be stable, with a specified confidence level. Such an approach has the potential to improve several steps in materials design by replacing the focus on high precision and accuracy with a more realistic and impactful focus on increased confidence in data and models.

Table 1: Probability of phase regions being stable at the mean eutectic location ($x_{Mg}$: 0.214, T: 1003K) is tabulated. Phases with no predicted probability of being stable are not included.

| Phase | Phase Probability |
|---|---|
| LIQUID | 0.22933 |
| FCC_A1 + LIQUID | 0.21467 |
| LAVES_C15 + LIQUID | 0.04133 |
| FCC_A1 + LAVES_C15 | 0.51467 |
| **PHASE TOTALS** | **1.00000** |

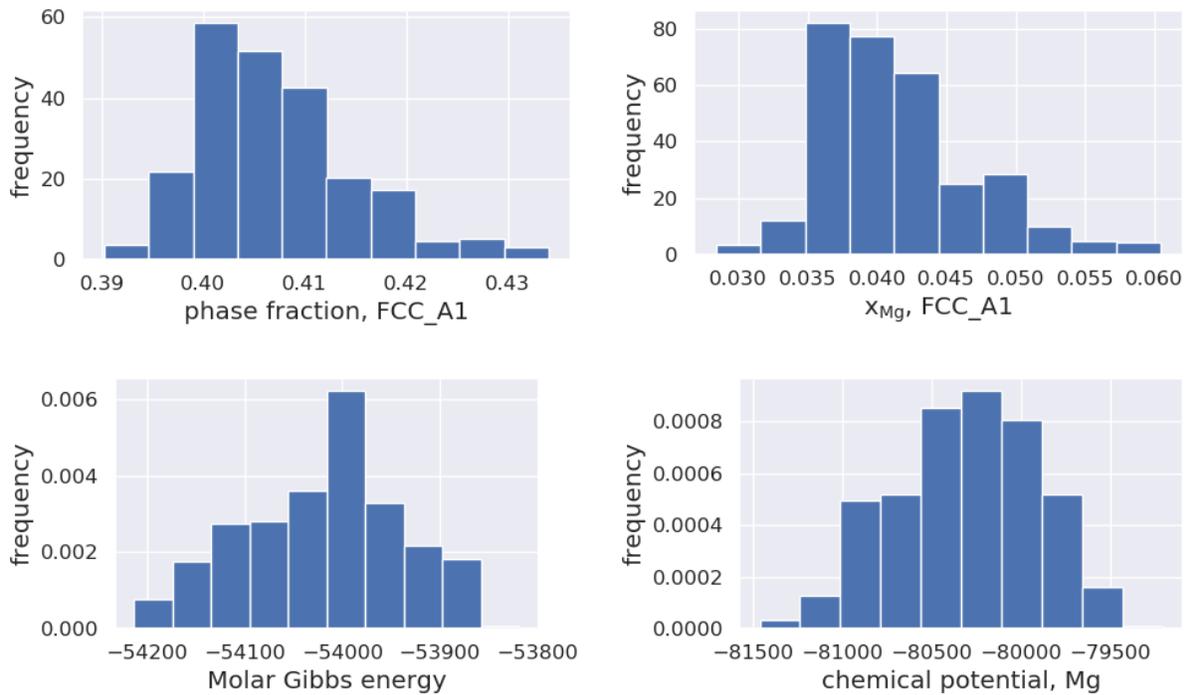

Figure 4: Probability distributions of (a) phase fraction, (b) phase composition of Mg, (c) Molar Gibbs energy, and (d) chemical potential of Mg are plotted for the FCC_A1 phase at the mean eutectoid location. Each histogram represents the properties of the 386 parameter vectors in which the FCC_A1+LAVES_C15 phase is stable.

Using the same set of equilibrium calculations employed to produce Fig. 2b, we plot in Fig. 5 the stability probabilities of each phase combination for $x_{Mg}$ between 0.0 and 0.35, and temperature between 950K and 1100K. While somewhat unwieldy, this representation provides significantly more information about the uncertainty of the binary phase diagram than the traditional approach where uncertainty intervals are plotted for each phase boundary, and presents a full, quantitative picture of the complex diversity of phase diagrams contained within the distribution of model parameter vectors from MCMC optimization.

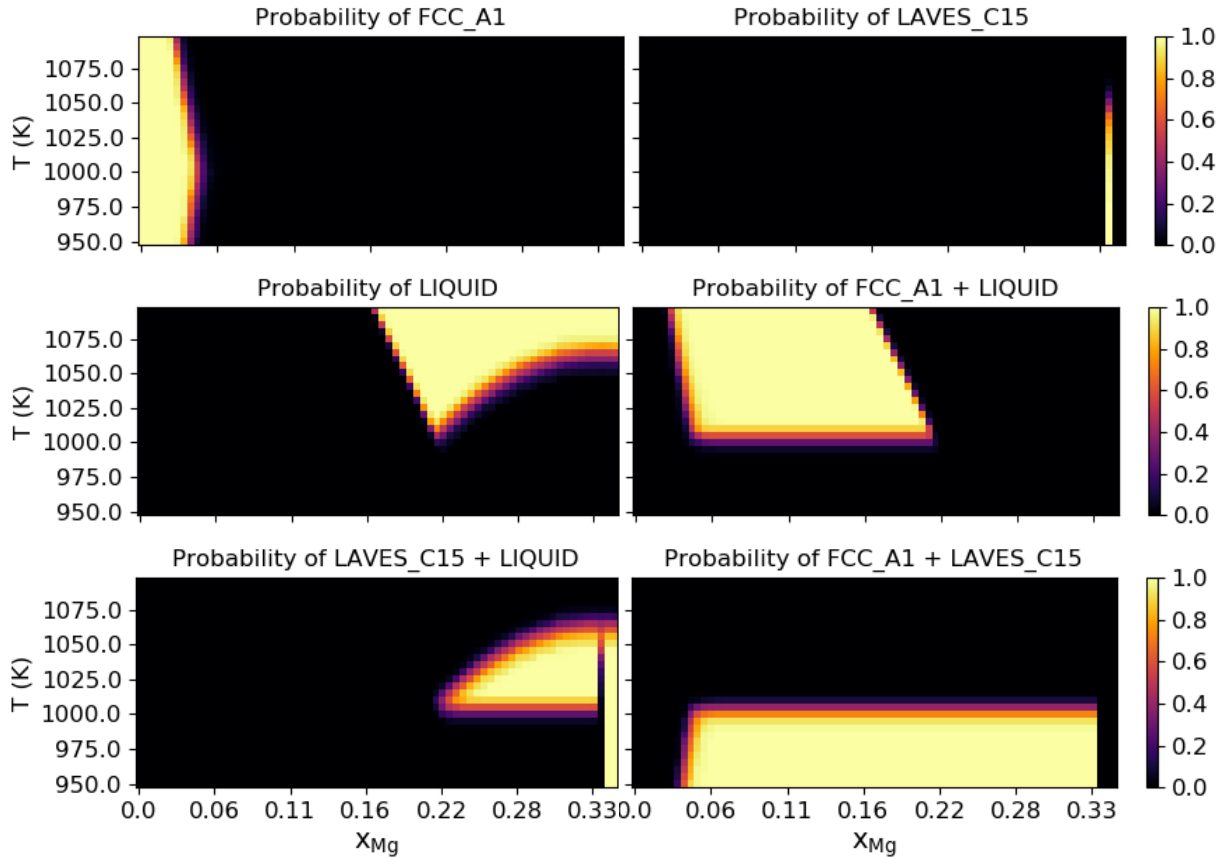

Figure 5: The probability of stability for (a) single phase FCC_A1, (b) single phase LAVES_C15, (c) single phase LIQUID, (d) FCC_A1+LIQUID, (e) LAVES_C15+LIQUID, and (f) FCC_A1+LAVES_C15 are plotted for $x_{Mg}$ between 0.0 and 0.35, and T between 950K and 1100K.

Lastly, we demonstrate the power of these approaches in quantifying the uncertainty of metastable phase diagram predictions. Metastable predictions are of great importance because they give insights into the non-equilibrium states that often occur during materials processing and define high-performance materials. Metastable regions of Gibbs energy space are not accessible to the vast majority of experiments, making it critical to convey the uncertainty in the predictions. Figure 6 presents the results of the Gibbs energy minimization for only the FCC_A1 and LIQUID phases. Figure 6a shows the predicted Gibbs energies and 95% uncertainty intervals (Bayesian credible intervals delineating the 2.5$^{th}$ and 97.5$^{th}$ percentiles of the Gibbs energies) for both phases. The large uncertainty interval in the Gibbs energy of the FCC_A1 phase means that the LIQUID phase can be marginally stable at this temperature. The superimposed phase diagram of Fig. 6b follows this conclusion. At the Cu-rich side of the phase diagram where FCC_A1 is in stable equilibrium with LIQUID, the phase boundaries are relatively less diffuse, with the FCC_A1 phase boundary being less certain than the liquid phase boundary. Figure 6b shows that the boundaries both become more diffuse as FCC_A1 becomes metastable with increasing Mg content. Even though the LIQUID phase is in stable equilibrium with the LAVES_C15, CUMG2, and HCP_A3 phases as the Mg composition increases, the uncertainty in the FCC_A1 Gibbs energy gives rise to the uncertain metastable phase boundary. Figure 6c presents the phase fractions and their 95% uncertainty interval versus temperature at $x_{Mg}$ = 0.2. Figure 6d gives the probability of stability for single phase LIQUID and indicates large uncertainties in metastable regimes (for T<1000K or $x_{Mg}$>0.2).

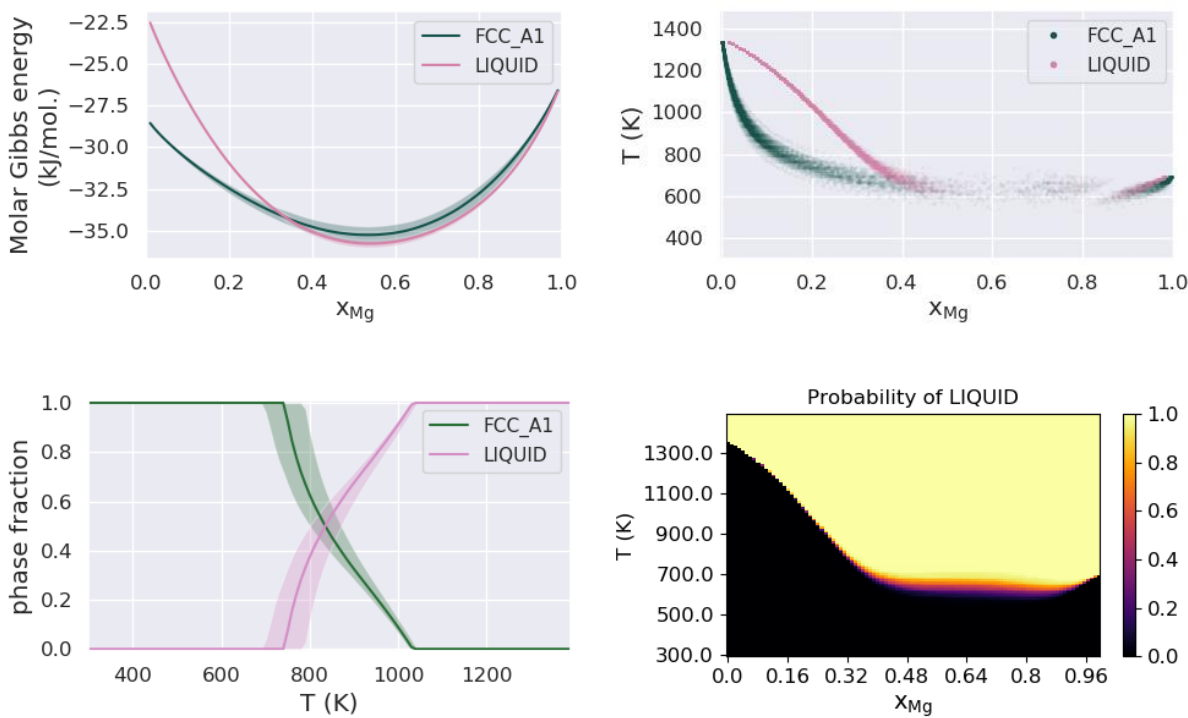

Figure 6: (a) Gibbs energies at 650K, (b) superimposed phase diagram, (c) phase fractions at $x_{Mg}$ = 0.2, and (d) single phase LIQUID stability probabilities are plotted for the FCC_A1/LIQUID metastable system.

# 4 Discussion and Conclusions

In this work, we present novel approaches to uncertainty quantification of all properties from CALPHAD modeling aimed at enhancing crucial materials design capabilities. We demonstrate the power of these methods by analyzing the Cu-Mg binary system via MCMC parameter samples generated using the open-source ESPEI Python package. These methods constitute an advancement in UQ of all properties from CALPHAD modeling beyond the representation of the uncertainty in phase boundaries. In short, the proposed methodology enables a detailed understanding of the potential for diverse and complex features to arise in the phase diagram, estimations of probability distributions of features such as invariant points, the probabilities of phase stability at any X-T-P point, and the phase-specific distributions of quantities of interest. Furthermore, these approaches are intuitively constructed from MCMC parameter samples, making them easy to implement alongside any CALPHAD optimization routine resulting in probability distributions of the model parameters.

In future work, we intend to demonstrate this approach through UQ tasks in systems with three or more components. In such systems, our method provides probabilities of phase stability without the need for visual representation – a principal difficulty in working with multi-component systems. Furthermore, we intend to integrate these tools into inverse materials design protocols where it may be desirable to, for example, identify an X-T-P location where a phase of interest is present within a specified range of phase fractions.

## Acknowledgements

N.H.P and M.S acknowledge financial support from awards 70NANB14H012 and 70NANB19H005 from U.S. Department of Commerce, National Institute of Standards and Technology as part of the Center for Hierarchical Materials Design (CHiMaD), and Laboratory Directed Research and Development (LDRD) funding from Argonne National Laboratory, provided by the Director, Office of Science, of the U.S. Department of Energy under Contract No. DE-AC02-06CH11357. R.O. acknowledges financial support from NASA's Science Mission Directorate and Space Technology Mission Directorate through the Game Changing Development program under Prime Contract #80NM0018D0004, and the Space Technology Office at the Jet Propulsion Laboratory, California Institute of Technology. B.B. and Z.K.L acknowledge financial support from the NASA Space Technology Research Fellowship under grant number 80NSSC18K1168, the National Research Trainee Fellowship under grant DGE-1449785 from National Science Foundation, and the Department of Energy under grant DE-FE003155.

## References

[1]     H. Lukas, S. G. Fries, and B. Sundman, *Computational thermodynamics: the Calphad method*. Cambridge university press, 2007.

[2]     G. B. Olson, "Computational design of hierarchically structured materials," *Science (80-. ).*, vol. 277, no. 5330, pp. 1237–1242, 1997.

[3]     M. Hillert, "The compound energy formalism," *J. Alloys Compd.*, vol. 320, no. 2, pp. 161–176, May 2001.

[4]     P. Honarmandi, T. C. Duong, S. F. Ghoreishi, D. Allaire, and R. Arroyave, "Bayesian Uncertainty Quantification and Information Fusion in CALPHAD-based Thermodynamic Modeling," *arXiv Prepr. arXiv1806.05769*, 2018.

[5]     R. A. Otis and Z.-K. Liu, "High-Throughput Thermodynamic Modeling and Uncertainty Quantification for ICME," *JOM*, vol. 69, no. 5, pp. 886–892, May 2017.

[6]     D. V Malakhov, "Confidence intervals of calculated phase boundaries," *Calphad*, vol. 21, no. 3, pp. 391–400, 1997.

[7]     E. Königsberger, "Improvement of excess parameters from thermodynamic and phase diagram data by a sequential Bayes algorithm," *Calphad*, vol. 15, no. 1, pp. 69–78, 1991.

[8]     W. Olbricht, N. D. Chatterjee, and K. Miller, "Bayes estimation: A novel approach to derivation of internally consistent thermodynamic data for minerals, their uncertainties, and correlations. Part I: Theory," *Phys. Chem. Miner.*, vol. 21, no. 1, pp. 36–49, May 1994.

[9]     N. D. Chatterjee, C. Miller, and W. Olbricht, "Bayes estimation: A novel approach to derivation of internally consistent thermodynamic data for minerals, their uncertainties, and correlations. Part II: Application," *Phys. Chem. Miner.*, vol. 21, no. 1–2, pp. 50–62, 1994.

[10]    N. D. Chatterjee, R. Krüger, G. Haller, and W. Olbricht, "The Bayesian approach to an internally consistent thermodynamic database: theory, database, and generation of phase diagrams,"

*Contrib. to Mineral. Petrol.*, vol. 133, no. 1, pp. 149–168, Oct. 1998.

[11] M. Stan and B. J. Reardon, "A Bayesian approach to evaluating the uncertainty of thermodynamic data and phase diagrams," *Calphad*, vol. 27, no. 3, pp. 319–323, 2003.

[12] T. C. Duong *et al.*, "Revisiting thermodynamics and kinetic diffusivities of uranium–niobium with Bayesian uncertainty analysis," *Calphad*, vol. 55, pp. 219–230, 2016.

[13] B. Bocklund, R. A. Otis, A. Egorov, A. Obaied, I. Roslyakova, and Z.-K. Liu, "ESPEI for efficient database development, modification and uncertainty quantification: application to the Cu-Mg system," *Prep.*

[14] Y. Sun, "Optimization Algorithms and Applications in Traffic Signal Control and Machine Learning," 2016.

[15] B. W. Silverman, *Density estimation for statistics and data analysis*. Routledge, 2018.

[16] R. A. Otis and Z.-K. Liu, "pycalphad: CALPHAD-based Computational Thermodynamics in Python," *J. Open Res. Softw.*, vol. 5, no. 1, 2017.

[17] A. Gelman, J. B. Carlin, H. S. Stern, D. B. Dunson, A. Vehtari, and D. B. Rubin, *Bayesian data analysis*. CRC press, 2013.

[18] N. H. Paulson, E. Jennings, and M. Stan, "Bayesian strategies for uncertainty quantification of the thermodynamic properties of materials," *arXiv Prepr. arXiv1809.07365*, 2018.

[19] J. Goodman and J. Weare, "Ensemble samplers with affine invariance," *Commun. Appl. Math. Comput. Sci.*, vol. 5, no. 1, pp. 65–80, 2010.

[20] D. Foreman-Mackey, D. W. Hogg, D. Lang, and J. Goodman, "emcee: the MCMC hammer," *Publ. Astron. Soc. Pacific*, vol. 125, no. 925, p. 306, 2013.